\documentclass[aps,prb,twocolumn,showpacs,superscriptaddress]{revtex4}
\usepackage{graphicx}
\usepackage{epsfig}
\usepackage{epstopdf}
\usepackage{color}
\usepackage{amsfonts}
\usepackage{amsthm}
\usepackage{amsmath}
\usepackage{amssymb}
\usepackage[latin1]{inputenc}
\usepackage[english]{babel}
\usepackage{pdfsync}
\usepackage{hyperref} 
\hypersetup{colorlinks=true}

\begin{document}


\def\beq{\begin{equation}}
\def\eeq{\end{equation}}
\def\be{\begin{equation}}
\def\ee{\end{equation}}

\def\iomn{i\omega_n}
\def\iom#1{i\omega_{#1}}
\def\c#1#2#3{#1_{#2 #3}}
\def\cdag#1#2#3{#1_{#2 #3}^{+}}
\def\epsk{\epsilon_{{\bf k}}}
\def\Ga{\Gamma_{\alpha}}
\def\Seff{S_{eff}}
\def\dinf{$d\rightarrow\infty\,$}
\def\T{\mbox{Tr}}
\def\t{\mbox{tr}}
\def\cG0{{\cal G}_0}
\def\cS{{\cal S}}
\def\divnum{\frac{1}{N_s}}
\def\vac{|\mbox{vac}\rangle}
\def\intR{\int_{-\infty}^{+\infty}}
\def\intb{\int_{0}^{\beta}}
\def\spinup{\uparrow}
\def\spindown{\downarrow}
\def\bra{\langle}
\def\ket{\rangle}

\def\ka{{\bf k}}
\def\vk{{\bf k}}
\def\vq{{\bf q}}
\def\vQ{{\bf Q}}
\def\vr{{\bf r}}
\def\q{{\bf q}}
\def\R{{\bf R}}
\def\kp{\bbox{k'}}
\def\a{\alpha}
\def\b{\beta}
\def\d{\delta}
\def\D{\Delta}
\def\e{\varepsilon}
\def\eps{\epsilon}
\def\ed{\epsilon_d}
\def\ef{\epsilon_f}
\def\g{\gamma}
\def\G{\Gamma}
\def\l{\lambda}
\def\L{\Lambda}
\def\o{\omega}
\def\ph{\varphi}
\def\s{\sigma}
\def\chib{\overline{\chi}}
\def\et{\widetilde{\epsilon}}
\def\hn{\hat{n}}
\def\hnu{\hat{n}_\uparrow}
\def\hnd{\hat{n}_\downarrow}

\def\hc{\mbox{h.c}}
\def\Im{\mbox{Im}}

\def\est{\varepsilon_F^*}
\def\v2o3{V$_2$O$_3$}
\def\uc2{$U_{c2}$}
\def\uc1{$U_{c1}$}


\def\bea{\begin{eqnarray}}
\def\eea{\end{eqnarray}}
\def \bal{\begin{align}}
\def \eal{\end{align}} 
\def\#{\!\!}
\def\@{\!\!\!\!}

\def\vi{{\bf i}}
\def\vj{{\bf j}}

\def\+{\dagger}


\def\up{\spinup}
\def\down{\spindown}


\def\'{\prime}
\def\"{{\prime\prime}}


\title{Hund's coupling key role in multi-orbital correlations}
\author{Luca de' Medici}
\affiliation{Laboratoire de Physique des Solides, UMR8502 CNRS-Université Paris-Sud, Orsay, France}

\begin{abstract}
We show how in multi-band materials, the Hund's coupling plays a crucial role in tuning the degree of electronic correlation. While in half-filled systems it enhances the correlations, in all other cases it pushes the boundary for the Mott transition at very high critical couplings.
Moreover in weakly-hybridized non-degenerate systems the Hund's coupling plays the role of band-decoupler, causing a change from a collective to an individual band behavior, due to the freezing of orbital fluctuations. In this situation the physics is strongly dependent on individual filling and electronic structure of each band, and orbital-selective Mott transitions (or even a cascade of such transitions) are to be expected. More generally a heavy differentiation in the actual degree of correlation of different bands arises and the system can show both weakly and strongly correlated electrons. 
\end{abstract}

\maketitle

\section{Introduction}\label{sec:intro}

Correlated materials are a continuous source of new and intriguing phenomena in condensed matter physics.
The realizable stoichiometric structures of compounds partially filling 3d, 4d, 5d, 4f or 5f orbitals are virtually endless and many of the explored ones have shown an extremely rich physics as a function of external parameters such as pressure, temperature, magnetic field and chemical substitutions.

This rich and varied physics is in many cases a direct outcome of electron-electron interactions and of the subtlety of the many-body physics that they yield.

The "correlation strength", is the relevance of these many-body effects for the conduction electrons.
These effects are indeed related to the interaction coupling strength in the conduction bands (but not always simply proportional to, as we will see in this paper), compared to their width (or more precisely the kinetic energy of the bare conduction electrons).
The coupling strength is maximal in the mentioned orbitals in that they have a reduced spatial extension, which maximizes the coulomb interaction within the shell, yet they are filled after the bigger s and p shells of higher principal quantum number, thanks to the aufbau "inversion" rule. Thus e.g. the 4s are filled before the 3d orbitals, and determine the atomic positions such that the overlap of the 3d orbitals, and the corresponding bandwidth, is rather small.

However these correlated orbitals are degenerate (5-fold for the d and 7-fold for the f) in isolated atoms and different behaviors arise in real solids depending on whether the crystal-field due to the surrounding atoms lifts or not this degeneracy. 

In materials where the degeneracy is totally lifted, only one band crosses the Fermi level and is thus partially filled, and a one orbital tight-binding model is often enough for a description of the low-energy physics. In this case the ratio between the inter-orbital interaction strength $U$ and the bandwidth $W$ is the control parameter of the many-body effects. 

When several correlated orbitals contribute to the conduction bands instead, more parameters come into play, and a richer many-body phenomenology is to be expected. Besides the more complicated band structure, the number of parameters needed to describe the electron-electron interaction is increased. Inter-orbital interactions are different from intra-orbital ones and Hund's rule, the tendency of electrons to distribute in different orbitals and in high-spin states, plays a role.

In a multi-band tight-binding low-energy model the Hamiltonian reads in general:
\be\label{eq:Ham}
H= \@ \sum_{ij, m m^\' \sigma}t^{mm^\'}_{ij} d^\dagger_{im\sigma}d_{jm^\'\sigma}+H_{int}
\ee
where $d^\dagger_{im\sigma}$ creates an electron with spin $\s$ in orbital $m$ on the site $i$ and $t^{mm^\'}_{ij}$ are the hopping amplitudes.
The local interaction vertex can be written as:
\bea\label{eq:H_int}
 H_{int}\@&=\@& U\sum_{i,m}
 n_{im\uparrow} n_{im\downarrow}+(U' -\frac{J}{2})\@\sum_{i,m>m' } n_{im} n_{im'} \\
 \#&-\#&\# J\@\sum_{i,m>m'}\@\# \left [ 2 {\bf S}_{im}\#  \cdot {\bf S}_{im'}\!
 +(d^\dagger_{im\uparrow}d^\dagger_{im\downarrow}d_{im'\uparrow}d_{im'\downarrow}\#+h.c.)\right]\#, \nonumber
 \eea
where $\bar\s\#=\#-\s$, $n_{im\s}\#=\# d^\dagger_{im\sigma}d_{im\sigma}$, $n_{im}\#=\# \sum_\s n_{im\s}$, ${\bf S}_{im}\#=\# \frac{1}{2}\sum_{\s\s'} d^\dagger_{im\sigma}{\boldsymbol  \tau}_{\s\s'} d_{im\sigma}$,  where ${\boldsymbol  \tau}_{\s\s'}$ is the Pauli matrices vector.
$U$ is the intra-orbital interaction, $U^\'$ is the inter-orbital one ($U^\'=U-2J$ holds from symmetry reasons on the matrix elements \cite{Castellani_V2O3,Fresard_Kotliar}) and $J$ is the Hund's exchange coupling.  
J can be seen then as a measure of the dependence of the interaction strength on the electrons occupying the same or different orbitals, and on their mutual spin alignment.

Among many tools for the investigation of correlated materials through the analysis of these low-energy models Slave-variable mean-fields (the most known being the Slave Bosons, SB)\cite{kotliar_ruckenstein,Coleman_Slave_bosons} and Dynamical Mean-Field Theory (DMFT)\cite{georges_RMP_dmft} have collected many successes. These are non-perturbative local mean-fields and have been successful in describing many aspects of correlated systems as band renormalization and mass enhancement, magnetic orders and the Mott transition.

The latter is the the transition between a paramagnetic metal and a paramagnetic insulator because of the localization of conduction electrons due to interactions. It is relevant to many materials\cite{imada_mit_review} and it has also been invoked to explain the physics of high-temperature superconductors\cite{Anderson_highTc_RVB}. 
The physics in the proximity of the Mott Transition has some well-defined features, that deserved the christening of a specific name: "Mottness".
These include strong quasiparticle mass renormalization and sizable spectral weight transfer from low-energy, coherent, to high-energy, incoherent features, among others.

The metallic phase in proximity to a Mott transition is a paradigm of  strongly correlated physics, and its analysis has been performed in detail in the one-band Hubbard model\cite{georges_RMP_dmft}.

At present there is less understanding of the Mott physics and in general of the effect of interactions in multi-band models.
Early studies\cite{Lu_gutz_multiorb,rozenberg_multiorb} have clarified that the Mott transition is found at any integer filling and in N-band degenerate systems with bandwidth $W$ and $SU(2N)$ symmetric interaction, i.e. $J=0$, the critical interaction strength $U_c/W$ grows with the degeneracy N and is maximal at half-filling.

The dependence of the Mott transition  critical coupling $U_c/W$ as a function of Hund's coupling $J$ was studied in Refs.\cite{Han_multiorb_Hund,Koga_multiorb_Hund,pruschke_Hund}. In all these studies $J$ is shown to \emph{reduce} the $U_c/W$ needed for the Mott transition, and thus to correlate the electrons. However these studies are restricted to half-filled systems and the generality of this statement for the Mott transitions at all integer fillings is questionable.

Recently, the role of Hund's coupling has been brought under the spotlight by the discovery of Iron-based superconductors\cite{Kamihara_pnictides1,Kamihara_pnictides2} in which this coupling within the Fe orbitals is believed to be sizable. Metallic iron is known to have a strong Hund's coupling, leading to ferromagnetism, and this is expected to be true also in the Fe atoms of these compounds. Moreover LDA+DMFT calculations have shown extreme sensitivity of some aspects of the physics of Fe-superconductors to the Hund's coupling strength\cite{Aichhorn_LaFeAsO,Aichhorn_FeSe, HauleShim_FeAs,Haule_FeAs}. In these materials 6 electrons occupy 5 correlated bands all crossing the Fermi level, mainly of Fe character. The application to this case of the present knowledge on Hund's coupling effect is not obvious.

One of the main results of this paper (section \ref{sec:deg}) is that for all integer fillings of a degenerate Hubbard model beside global half-filling the Hund's coupling actually \emph{increases} the critical U needed for the Mott transition, bringing it to very large values for reasonable coupling strength, contrary to the previously studied half-filled cases.

Many other aspects of the physics of Fe superconductors are still subject to a debate in the scientific community, and many competing scenarios have been put forward, without reaching consensus thus far. The possibility of these materials being in or in proximity to an Orbital Selective Mott Phase in which some localized electrons coexist with itinerant ones has been invoked by several groups\cite{Wu_FeAs_OSMT,Kou_OSMT_pnictides,demedici_Genesis,Hackl_Vojta_OSMT_pnictides}. The mechanism possibly leading to an Orbital-Selective Mott Transition\cite{Anisimov_OSMT} (OSMT) in these materials has been first isolated in Ref. \cite{demedici_3bandOSMT} and shown to be related to the strong Hund's coupling.

The other main result (section \ref{sec:nondeg}) of this paper is the highlighting of Hund's coupling as the main factor responsible in general for the arising of OSMT in real non-degenerate (for the presence of a crystal-field splitting or different band dispersions) weakly hybridized systems. Overriding or accompanying bandwidth differences and degeneracies, Hund's coupling is seen to act as a \emph{band-decoupler}, through the suppression of orbital fluctuations.  Then the correlation character of each band is determined by its individual filling and structure. This mechanism also applies to the well known anomalous superconductor $Sr_2RuO_4$ as has been recently outlined in Ref.\cite{Mravlje_SrRuO4_coherence}.

\section{Models and Techniques}

We will solve the model eq.(\ref{eq:Ham}) with Slave-spin mean-field (SSMF)\cite{demedici_Slave-spins,Hassan_CSSMF} and DMFT\cite{georges_RMP_dmft}. 

SSMF is a computationally fast slave-variable mean field particularly suited for studying multi-band models. It is extremely useful for tracing entire phase diagrams and yields qualitatively accurate and semi-quantitative results, in line with the performance of usual slave-bosons. But unlike the latter, which become quickly untractable in the multi-band case, SSMF limit the proliferation of auxiliary variables and can thus tackle i.e. a five band model and trace a phase diagram as a function of several parameters.
As all other slave-variable methods, at the mean-field level SSMF approximates the model with a quasi-particle effective non-interacting model whose parameters are self-consistently determined. Solving a local problem for the slave variable takes into account the effect of interactions and yields the quasiparticle effective parameters, i.e. renormalization of the crystal-field and of the electron mass. The latter is expressed through the quasiparticle renormalization factor $Z_m$, which ranges from 1 (non-interacting system) to 0 and whose vanishing signals a Mott transition 

For supporting the SSMF results, and reporting aimed and quantitatively accurate results, we make use of the well established DMFT which is also heavier to run. DMFT assumes the locality of the self-energy and maps the (multi-orbital) lattice problem onto a (multi-orbital) quantum impurity problem subject to a self-consistency condition. The impurity problem has to be solved numerically, and we use Lanczos exact diagonalization\cite{Caffarel_Krauth} as impurity solver.

In all cases we consider non-hybridized bands, i.e. $t^{mm^\'}_{ij}=t^m_{ij}\d_{mm^\'}$. We then discuss how our analysis is affected by a finite (small) hybridization.

The two methods being local mean-fields, in absence of hybridization the k-dependence enters the problem only through each band dispersion. Sums over momenta can thus be replaced by integrals over energy weighted by density of states (DOS) $D(\eps)$, which we assume to be semicircular throughout the paper, i.e.:
\be
D(\eps)=\frac{2}{\pi D}\sqrt{1-(\eps/D)^2},
\ee
where $D=W/2$ is the half-bandwidth.

Then a model is fully specified by each band's width and position, the latter being set by the crystal-field splitting, i.e. by the set of on-site energy levels $\eps_m\equiv t^m_{ii}$.

We solve the model  in different cases: N-fold degenerate model in section \ref{sec:deg}, 3-band model with same bandwith and trigonal crystal-field splitting, and with three different bandwidths but without crystal field in section \ref{sec:nondeg}. 

We will assume zero temperature throughout the paper.

\section{Hunds coupling and the Mott transition in degenerate systems}\label{sec:deg}

We first study the N-fold degenerate Hubbard model at all possible integer fillings. All N bands have $D=1$ and no crystal-field splitting is present, i.e. $\eps_m=0$.

\begin{figure}[htbp]
\begin{center}
\includegraphics[width=8cm]{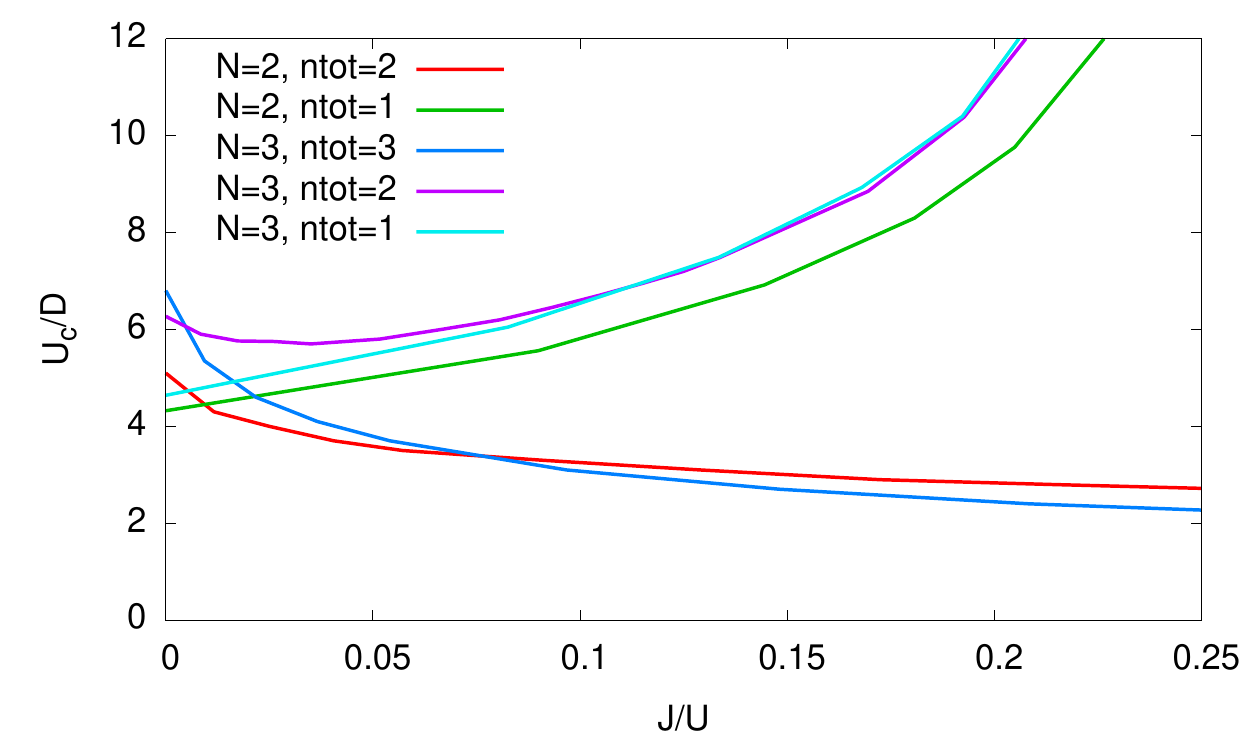}
\end{center}
\caption{Slave-Spin Mean-Field metal-Mott insulator critical coupling $U_c$ for the N-band degenerate Hubbard model as a function of the ratio $J/U$. Only fillings such that $n< N$ are shown here. The ones for $n>N$ give identical results due to particle-hole symmetry. For half-filled cases (2 electrons in 2 bands and 3 electrons in 3 bands) the Hunds coupling correlates the system and reduces the critical $U_c/D$. For all other fillings on the contrary $U_c/D$ is strongly increased.}
\label{fig:Uc_vs_ratioJU}
\end{figure}

In fig.\ref{fig:Uc_vs_ratioJU} we show the results of the SSMF. 
At all integer filling $n$ we find, as known, that for increasing U at fixed $J/U$ the renormalization factor $Z_m$ - identical for all bands - decreases from unity, reaching zero beyond a critical $U_c/D$ and thus signaling a Mott transition.
At J=0 our analysis confirms that $U_c$ is maximum for half-filling and that for a given filling $n/N$, Uc increases with the degeneracy N (see Ref. \cite{Lu_gutz_multiorb}).

We then study the dependence of the $U_c/D$ on the value of J.
At half-filling ($n=2$ electrons in $N=2$ bands, or  $n=3$ electrons in $N=3$ bands in figure \ref{fig:Uc_vs_ratioJU}) we find that increasing $J/U$ reduces the value of $U_c$, as reported in previous works\cite{Han_multiorb_Hund,Koga_multiorb_Hund,pruschke_Hund, Werner_Hund}.

But surprisingly, for any other filling $U_c$ \emph{is rather increased by Hund's coupling}.

\begin{figure}[htbp]
\begin{center}
\includegraphics[width=8cm]{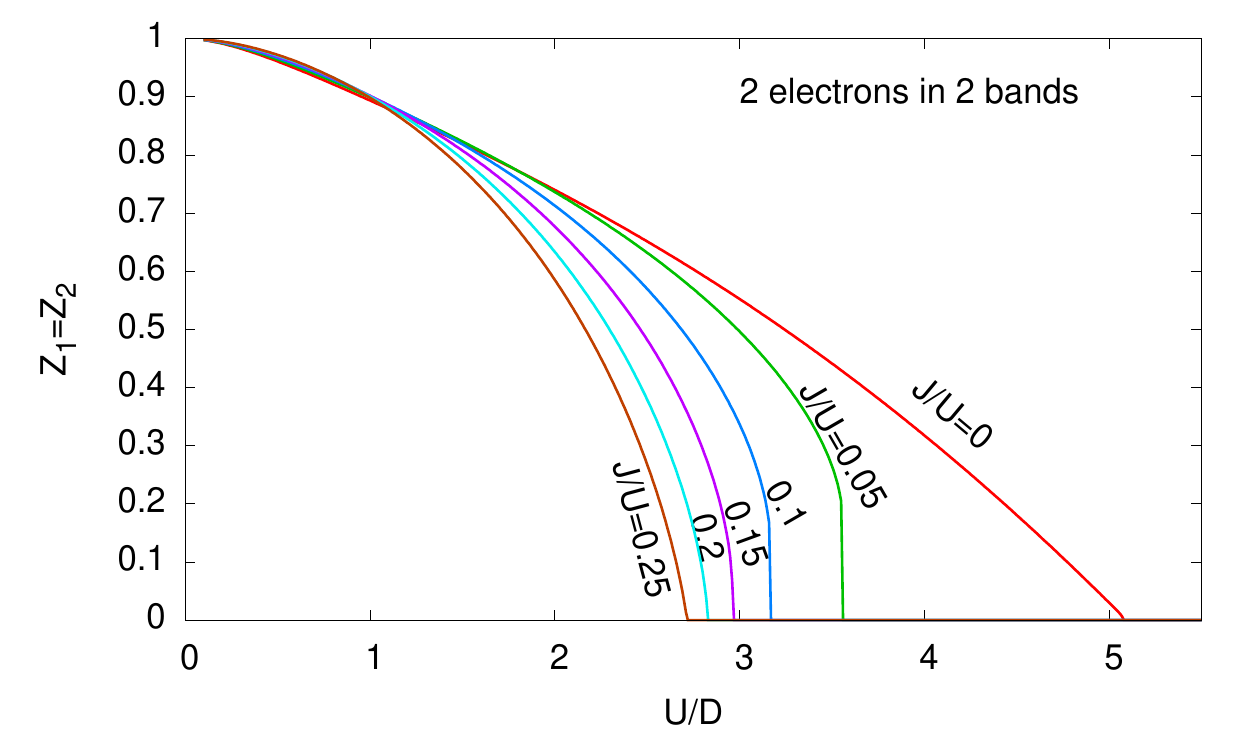}
\includegraphics[width=8cm]{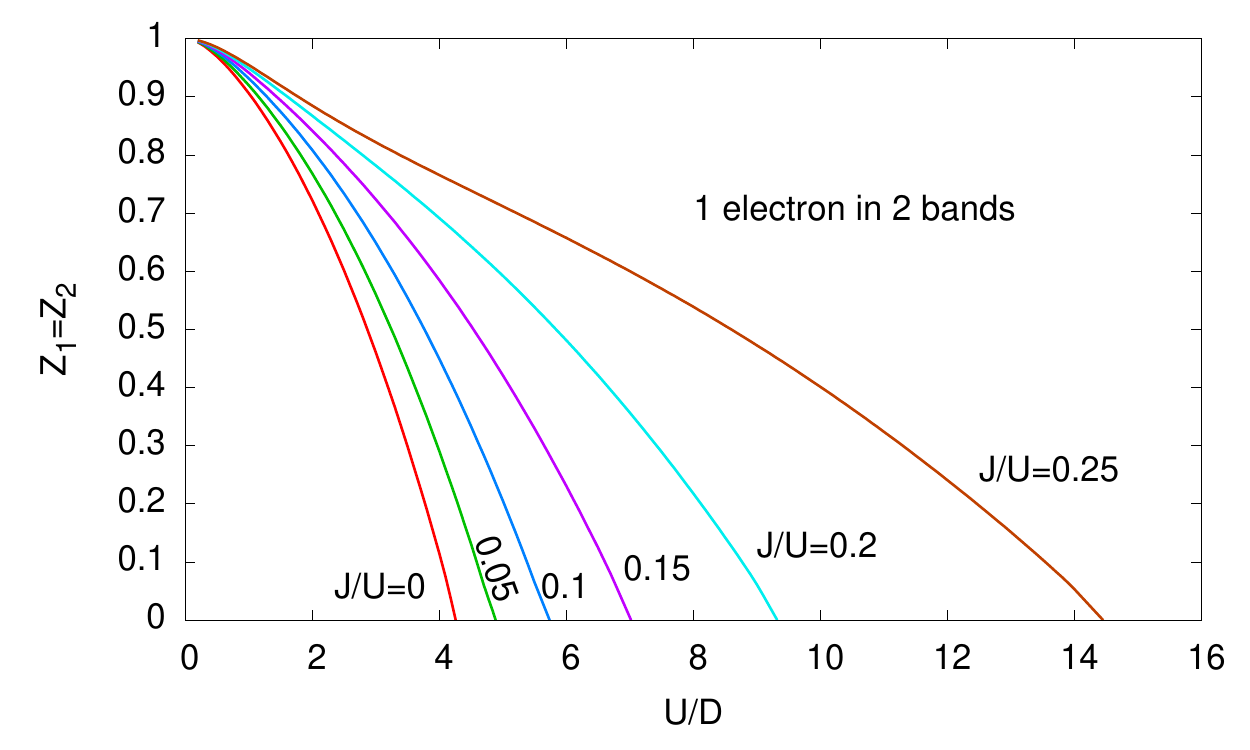}
\end{center}
\caption{Quasiparticle renormalization factor $Z$ for the 2-band Hubbard model as function of the interaction strength U and for different values Hund's coupling J. Upper panel: half-filled case. Lower panel: quarter-filled case. The effect of the Hund's coupling is opposite in the two cases.}
\label{fig:Z_vs_U}
\end{figure}

We see this in detail for the 2-band case in figure \ref{fig:Z_vs_U}. Upon going from $J=0$ to a typical value $J/U=0.25$, the critical interaction strength is reduced to a half of its original value at half-filling, but it increases to more than 3 times the $J=0$ value, for the quarter-filled case. 

This result is general and does not depend on the number of bands.
In the three-band model the reduction of $U_c$ at half-filling is even stronger than in the two-band case, while the behavior of the system with 1 electron follows the trend of the analogous case in the two-band system.
Interestingly a slightly different behavior can be noticed in the case of 2 electrons in 3 bands. Here the onset of J induces initially a reduction of $U_c$, which quickly starts increasing again and follows the trend of the other non-half-filled cases, with the Mott transition pushed at very strong couplings.

Hints of this physics can be seen in Ref. \cite{Werner_3band} where the opposite trend of the different ($n=1,2,3$) Mott insulating lobes of the phase diagram as a function of J was reported.

In order to understand these opposite trends we will follow an argument based on the analysis of the atomic limit\cite{gunnarsson_fullerenes,Han_multiorb_Hund,Werner_3band}.

Indeed an estimate of the $U_c$ can be obtained following Mott's original image for the single-band case: at strong coupling the local spectral function of a Mott insulator shows features based on the atomic excitations (two delta-functions at a distance U from one another),  broadened by the delocalization energy which can in principle be calculated perturbatively in $t/U$. In practice these broadened features (called the 'Hubbard bands") have a width in energy of the order of the bare bandwidth W. Thus an estimate of the Mott gap is the distance of the borders of these two features, i.e. $\D=U-W$. By the same token an estimate of the critical U for the Mott transition is given by the vanishing of the gap, i.e. $U_c/W=1$, which is not far from the $U_{c1}/W\simeq 1.2$ obtained by full-fledged DMFT.

The full many-body physics is more complicated\footnote{Indeed the DMFT picture of the Mott transition is more articulated and around the actual transition one finds a zone where a metallic and an insulating solutions coexists. Then the $U_{c1}$ where the gap closes and the insulating solution is lost is different from the $U_{c2}$ where the metallic solution disappears because the quasiparticle weight vanishes. This latter quantity is identified with the renormalization factor $Z_m$ calculated in SSMF. Although $U_{c1}$ and $U_{c2}$ may scale differently with the number of bands\cite{Florens_Multiorb}, it can be safely assumed that they both increase or they both decrease as a function of physical parameters.}, and in this estimates one should for example take into account the dependence of the delocalization energy, and thus of the width of the Hubbard bands,  on the physical parameters. In practice, at least in the Hubbard model this is seen to be a minor effect and only very close to the transition one sees sizable deviations from  $\D=U-W$.

\begin{figure}[htbp]
\begin{center}
\includegraphics[width=8.5cm]{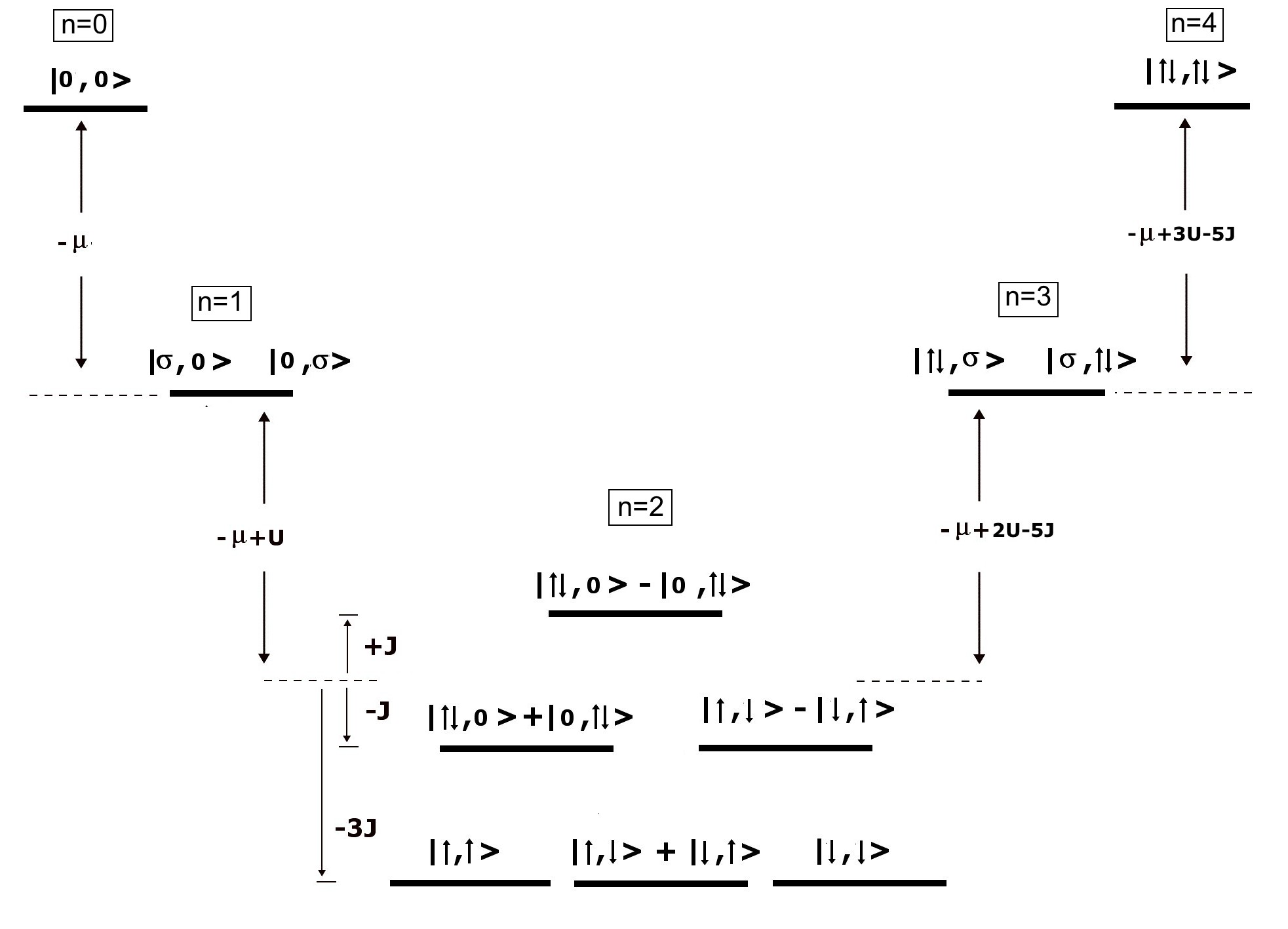}
\end{center}
\caption{Spectrum of a 2-orbital atom with Hamiltonian eq. \ref{eq:H_int}. The different sectors position in energy is as depicted for an half-filled system, for which the chemical potential can be obtained by particle-hole symmetry arguments and is $\mu=(3U-5J)/2$ and the ground state lies in the $n=2$ sector. In general the distance of this with the lowest lying state in the $n=3$ sector is $E^{3}_0-E^{2}_0=-\mu+2U-2J$ whereas the distance with the lowest lying state in the $n=1$ sector is $E^{1}_0-E^{2}_0=\mu-U+3J$. The atomic gap is then $\D=U+J$ and thus is enhanced by J.}
\label{fig:atomic_half}
\end{figure}

Transposing this argument to the multi-orbital case we thus want to obtain the scaling with Hund's $J$ of the Mott gap and of the $U_c$ for the Mott transition in the multi-band Hubbard model.
In Fig. \ref{fig:atomic_half} is depicted the atomic excitation spectrum for a 2-orbital atom. A change in the atomic spectrum will reflect directly on the distance between the Hubbard bands and then on the Mott gap.
The 16 states are divided in sectors with the same number of particles, whose distance is set by U. Instead within every sector, for a given particle number, the states are split by J.

The global filling of the system is set by the chemical potential which shift the relative position of the different sectors. When the system is half-filled $\mu$ is such that the ground state is in the 2-particle sector. For any finite $J$ this is 3-fold degenerate and is the triplet of $S=1$ states. The gap in the atomic excitation spectrum is obtained summing the energy distance between the ground state and the lowest lying levels ($E_0$) in the $n+1$ and $n-1$ sectors, i.e.:
\be
\D^{at}=E^{n+1}_0-E^{n}_0+E^{n-1}_0-E^{n}_0=E^{n+1}_0+E^{n-1}_0-2E^{n}_0.
\ee
This quantity is readily evaluated, yielding:
\be
\D^{at}=U+J.
\ee
It is independent of $\mu$ and shows that the gap is enlarged by J. In the Mott insulator then, if we assume that the Hubbard bands are broadened versions of the atomic excitations the gap will also scale linearly in J, and so will $U_c/D$ (with opposite sign, because a larger gap will close, when the Hubbard bands overlap, at a smaller U).

\begin{figure}[htbp]
\begin{center}
\includegraphics[width=8cm]{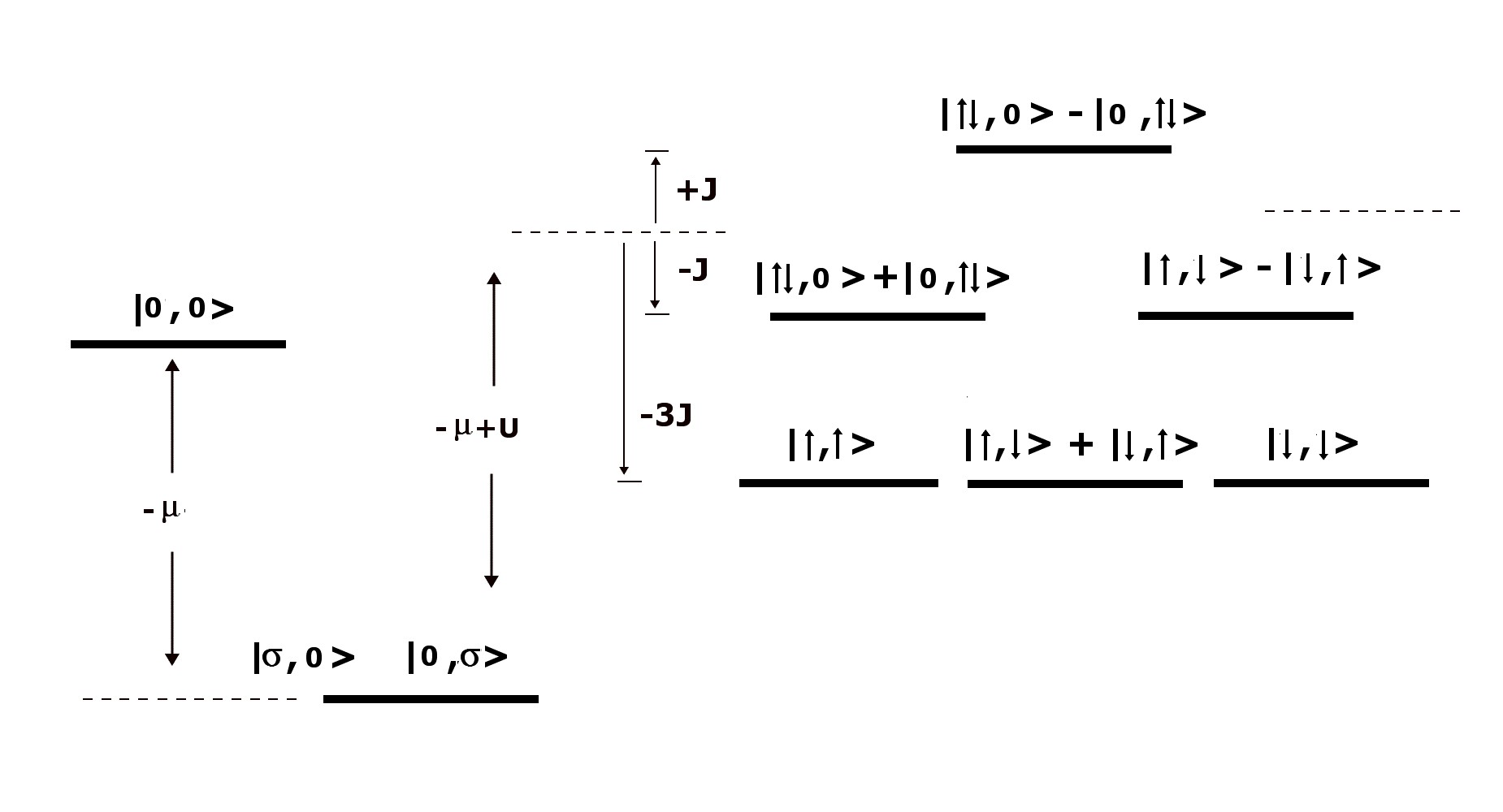}
\end{center}
\caption{Same as in Fig. \ref{fig:atomic_half} but sectors position in energy is depicted here for the system quarter-filled. The distance with the $n=2$ sector is $E^{2}_0-E^{1}_0=-\mu+U-3J$ whereas the distance with the $n=0$ sector is $E^{0}_0-E^{1}_0=\mu$. The atomic gap is then $\D=U-3J$ and is then reduced by J.}
\label{fig:atomic_quarter}
\end{figure}

Let's now replicate this argument for the quarter-filled case.
In the corresponding atomic spectrum (Fig. \ref{fig:atomic_quarter}) the lowest lying sector is the one with $n=1$.
The atomic gap is now:
\be
\D^{at}=U-3J.
\ee
which is reduced by J. Thus the $U_c$ for the Mott insulator with 1 particle (or 3) in 2 bands will get enhanced by J.

The same argument applies in general for any number N of bands, to show that the Mott gap gets reduced in all cases but at half-filling. It is easy to calculate that in general:
\be
\D^{at}_n=\begin{cases}
    U-3J, & \; \forall n\neq N \quad \text{(off half-filling)},\\
    U+(N-1)J,  & \; n=N  \quad \text{(half-filling)}.
  \end{cases}
 \ee

The rationale behind this is the following. The energy of the high-spin state in each sector is lowered by the exchange among the electrons which is proportional to the number of possible pairs of aligned spare spins. The latter is maximum in the half-filled sector where there is one spare electron in every orbital. Moving away from half-filling it decreases quadratically, because the number of the possible pairs is $n_s(n_s-1)/2$, where $n_s$ is the number of spare electrons, which decreases while the number of either empty or fully occupied orbitals increases.

Thus the gap in a half-filled Mott Insulator is always enhanced by J, being enhanced the distance between the ground state and the lowest state in the neighboring sectors. 

In the cases of a Mott insulator with a non half-filled ground state (say without loss of generality lying in the sector with $n<N$ particles) this has a gain in energy due to J which is smaller than the one for the high-spin state of the neighboring sector (with $n+1$ particles) closer to (or at) half-filling and bigger than the one with $n-1$ particles due to the mentioned quadratic dependence. The latter is also responsible for the fact that the subsequent enhancement of $E_{n-1}-E_{n}$ with J is less than the reduction of $E_{n+1}-E_n$. The overall gap is then reduced. 

The same obviously applies for the particle-hole symmetric situation for $n>N$.

The estimate of the dependence of $U_c$ on J, following these arguments is:
\be\label{eq:asynt}
\d U_c(n)\propto\begin{cases}
    3J, & \forall n\neq N \quad \text{(off half-filling)},\\
    -(N-1)J,  & n=N  \quad \text{(half-filling)}.
  \end{cases}
 \ee

\begin{figure}[htbp]
\begin{center}
\includegraphics[width=8cm]{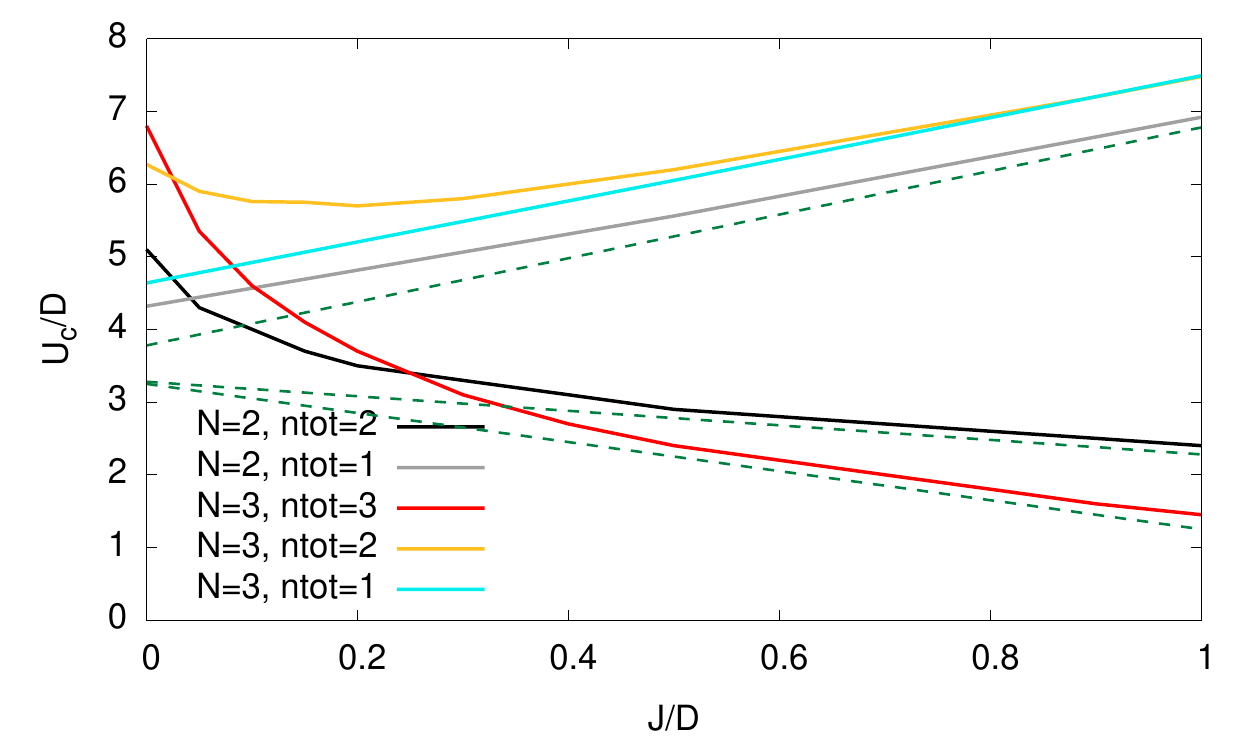}
\end{center}
\caption{(Color online)  Same phase diagram as in Fig. \ref{fig:Uc_vs_ratioJU}, plotted here as a function of J.  The dashed lines indicate the strong J behavior calculated analytically in the atomic limit, eq. (\ref{eq:asynt}) (from top to bottom $\d U_c\propto 3J, -J, -2J$).}
\label{fig:Uc_vs_J}
\end{figure}

In Fig. \ref{fig:Uc_vs_J} we plot the SSMF phase diagram as a function of J, and we compare the results with these asintotic behaviors based on the atomic limit eq. (\ref{eq:asynt}).  It is clear that at strong J the atomic limit captures perfectly the dependence of $U_c$.

However departures from these analytic results are found at weak J.
The previous argument basis indeed, is that J is large enough for only the high-spin state to be taken into account in each sector. This assumption is not justified at weak J and orbital fluctuations become important when taking into account the hopping from site to site.
These are maximal at $J=0$ and the consequent gain in kinetic energy is the cause for the enhancement of $U_c$ with the band degeneracy\cite{Florens_Multiorb}. The onset of J suppresses these fluctuations rapidly and we thus observe a quick reduction of $U_c$ in the half-filled systems, where the orbital fluctuations are maximal. Moving away from half-filling this effect is controlled by $n/N$ (for $n<N$, and by the particle-hole symmetric $(N-n)/N$ for $n>N$) and it is already negligible for the system with 1 particle in 3 bands.
 The reduction of orbital correlations is in fact a main effect of J and we will see in section \ref{sec:nondeg} its key importance in decoupling the orbitals.

It is worth stressing that these conclusions are drawn for degenerate non-hybridized bands but are still valid for small hybridizations $V_{mm^\'}$ and crystal-fields $\D_{mm^\'}=\eps_m-\eps_{m^\'}$. Indeed both hybridization and crystal-fields work in general "against" J (the hybridization favors singlet states, while the crystal-field favors orbital disproportionation, both thus favor low-spin states\cite{Werner_Hund}), but are non-singular perturbations. 

Thus if $J\gg V_{mm^\'}, \D_{mm^\'}$ we expect these results to hold in general.

More specifically one can show that their effect on the high-spin states in the atomic limit is null in the half-filled sector and lowers the energy in sectors away from half-filling. The detail of this energy gain depends in general on the specific structure of the  matrices $V_{mm^\'}, \D_{mm^\'}$ and its analysis is beyond the scope of this paper but it can be easily shown for N=2 and 3 that they always reduce the effect of J on the gap between the atomic sectors and will thus reduce the effect of J on $U_c$.
Some specific models with finite crystal-fields are studied in section \ref{sec:nondeg}.

SSMF and atomic limit considerations are somewhat complementary points of view on the Mott Transition and they draw a quite reliable picture, but one may want to benchmark these results with a more quantitative and accurate method as DMFT.

\begin{figure}[htbp]
\begin{center}
\includegraphics[width=8cm]{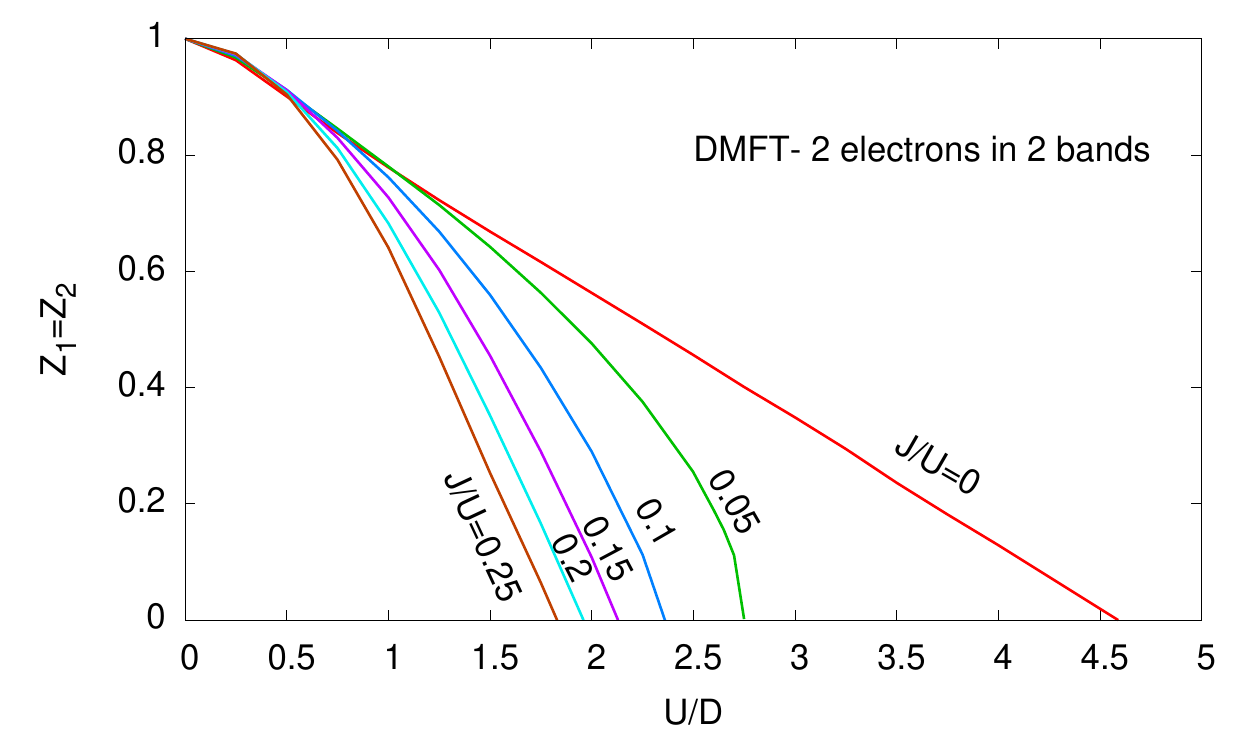}
\includegraphics[width=8cm]{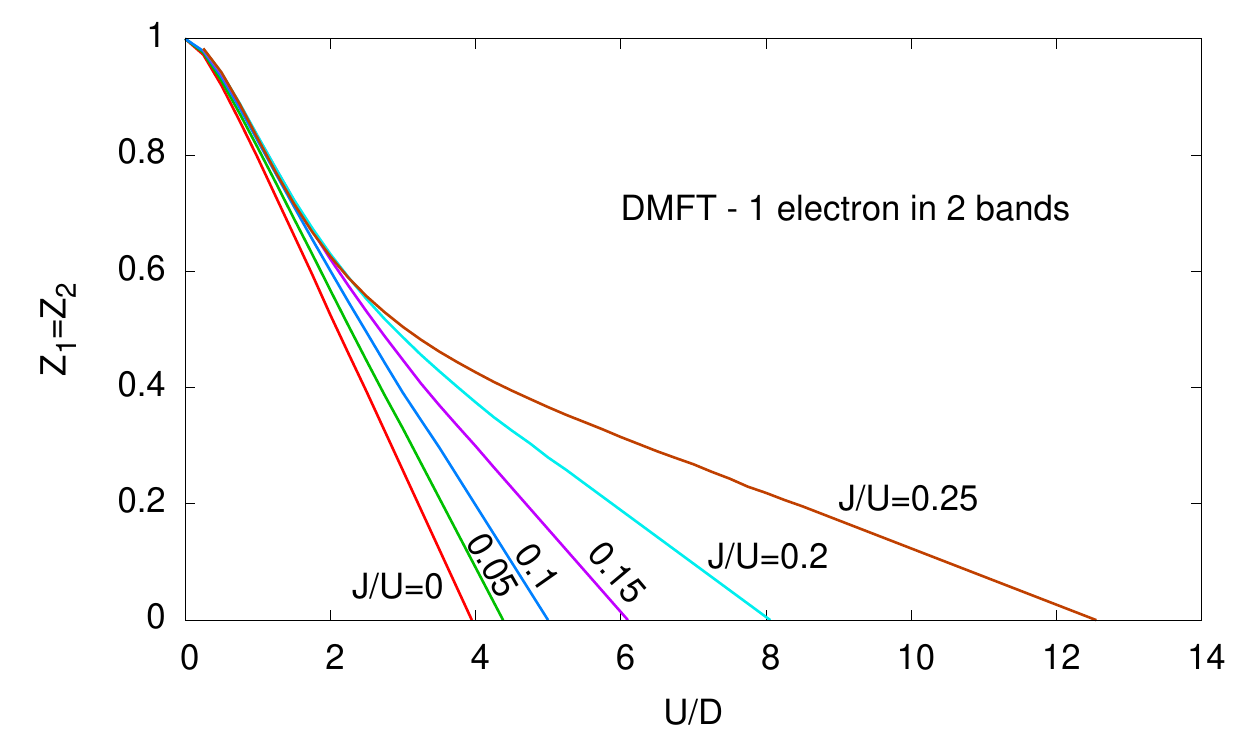}
\includegraphics[width=8cm]{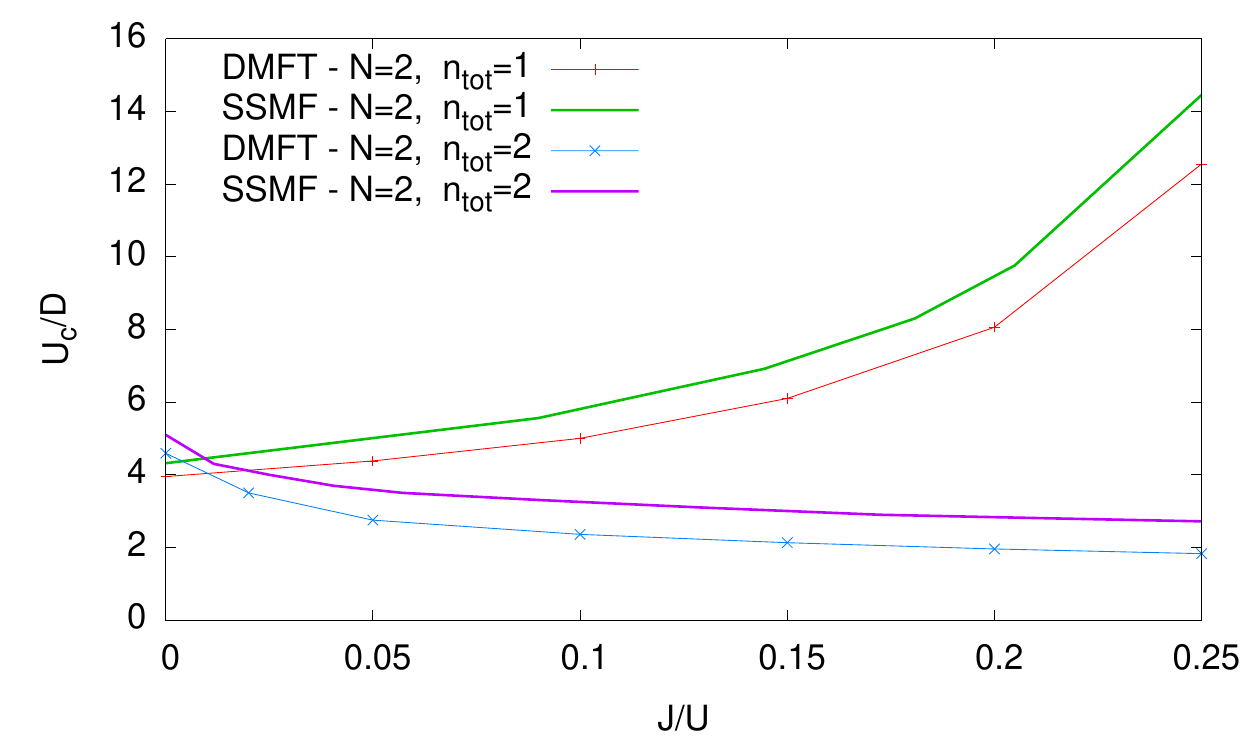}
\end{center}
\caption{DMFT results for 2-band degenerate Hubbard model for the Mott transition with $n=2$ electrons (upper panel), and $n=1$ (middle panel) which fully confirm SSMF results. Lower panel: phase diagram comparison between Slave Spin Mean-Field and Dynamical Mean-Field Theory. The slight overestimation of $U_c$ is a known issue of slave-variable mean-fields, common i.e. to slave bosons.}
\label{fig:DMFT_deg}
\end{figure}

We plot the DMFT results for the $N=2$ degenerate model in Fig. \ref{fig:DMFT_deg}. DMFT fully confirms the SSMF scenario and the reliability of the SSMF method. The transition boundaries are slightly shifted by DMFT as SSMF, like slave bosons, is known to overestimate $U_c$. 

DMFT can also yield a much deeper insight into the physics of the systems through the analysis of dynamical quantities as i.e. the local self-energies $\Sigma_m(\iomn)$ and spectral functions $\rho_m(\omega)=-1/\pi Im G_m(\omega+i0^+)$, where $G_m$ is the Green Function for band $m$.

\begin{figure}[htbp]
\begin{center}
\includegraphics[width=8cm]{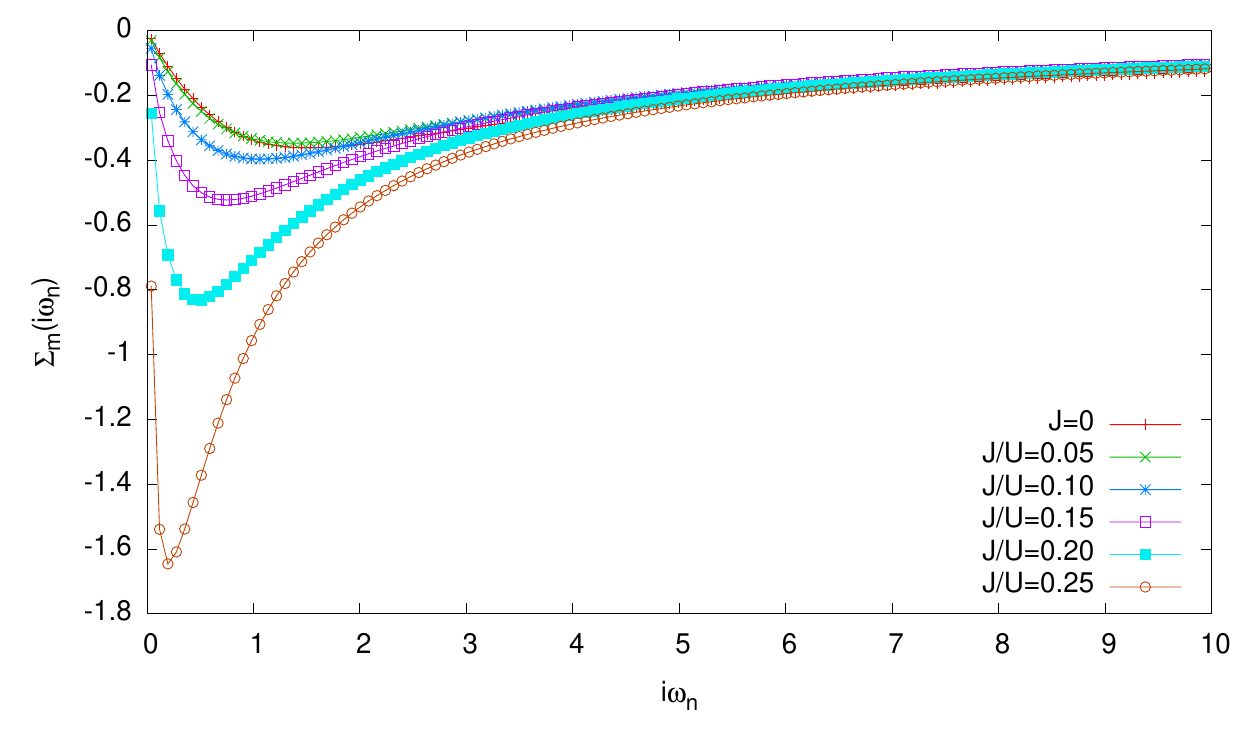}
\includegraphics[width=8cm]{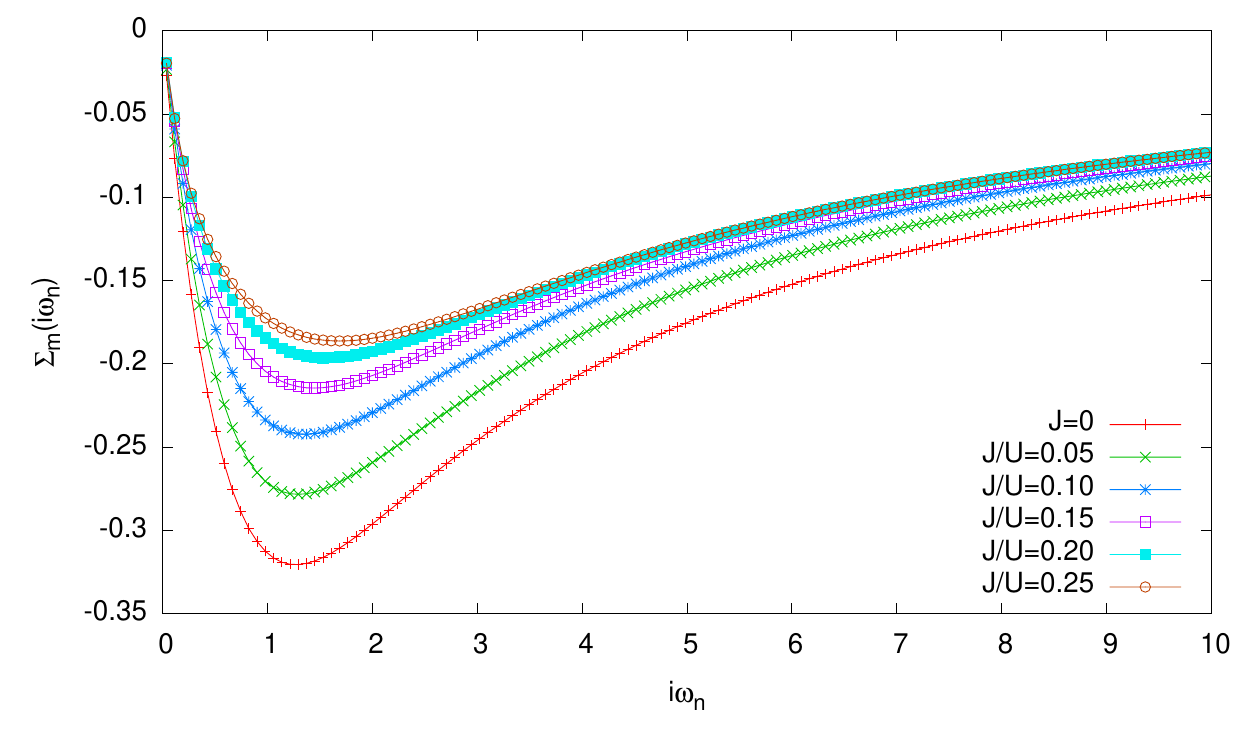}
\includegraphics[width=8cm]{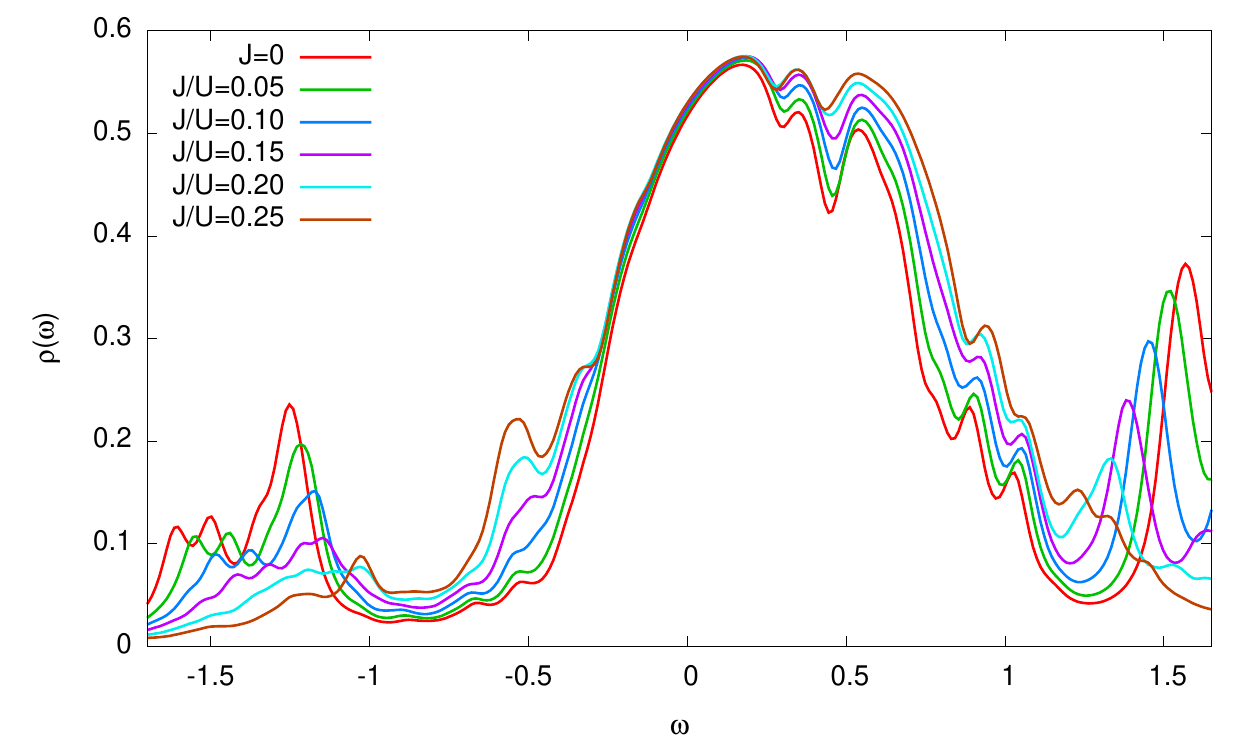}
\end{center}
\caption{Self-energies of the metallic phase calculated in DMFT for $U/D=1.75$ as a function of J. Upper panel: 2 electrons in 2 bands, the self-energy grows with J. Central panel: 1 electrons in 2 bands, the self-energy diminishes with J. Lower Panel: spectral functions for the quarter-filled case at low energy. The features of the Hubbard bands closest to the broad low-energy feature (the left-most and right-most in the plotted energy window) move closer as J is increased, as predicted by the atomic limit arguments.}
\label{fig:DMFT_self-energies}
\end{figure}

We plot in Fig. \ref{fig:DMFT_self-energies} the self-energies at $U/D=1.75$ as a function of J for the half-filled and quarter filled cases.  As expected J is seen to enhance the self-energy in the half-filled case, thus correlating the system, while it lowers the self-energy in the quarter-filled case and it decorrelates the system.

In Fig.\ref{fig:DMFT_self-energies} we plot the spectral functions corresponding to the quarter-filled 2-band Hubbard model in the metallic phase. The Hubbard bands are very structured reflecting the multiple energy scales present already in the atomic Hamiltonian\cite{Kotliar_Kajueter_Multiorb}.
In these bands, the features which are closer to the central peak (that is due to the coherent quasi-particle excitations which disappear at the Mott transition) are the precursors of the lowest atomic excitations. As predicted by the atomic limit arguments these features move closer to the central peak with J.

We can thus unambiguously draw the conclusion that Hund's coupling works against the Mott transition in all cases but the half-filled one.  \emph{Caveat} to this statements are only the reentrant behaviour of $U_c$ (at small J, where the suppression of orbital fluctuations overruns the shifting of the Hubbard bands) that can be found for integer fillings close to half, and the effect of strong hybridization and/or crystal field splitting in the regime $V_{mm^\'} \; \text{or} \; \D_{mm^\'} > J$.

Also, moving away from the Mott transition boundary, a subtle interplay of  this Mott atomic-like physics with the Kondo screening characterizes the correlated metallic phase. 
In the 2-band Hubbard model these two aspects work together. In the half-filled case the Kondo temperature is known to be reduced by J \cite{pruschke_Hund, Nevidomskyy_KondoHund}, while for the quarter-filled case a modest increase of it has been reported\cite{pruschke_Hund}, all in line with our results.

For the 3-band model the situation is expected to be analogous both for the half-filled and the extremal $n=1$ ($n=5$) cases. In the $n=2$ ($n=4$) case instead, the relation between the Kondo temperature and the $U_c$ dependence on $J$ is probably less straightforward, in view of the reentrant transition boundary visible in Fig.\ref{fig:Uc_vs_J} (also in Ref. \cite{Werner_SpinFreezing} a "spin freezing transition" is reported to happen before the Mott transition)  and this will be the subject of a future publication.

\section{Hund's coupling as a band-decoupler: Orbital-Selective Mott Transitions}\label{sec:nondeg}

Thus far we have only dealt with the fully degenerate Hubbard Model, where all bands are identical, and we have focused on the influence of the Hund's coupling on the behavior of the system as a whole.

We will now allow for differences between the bands and study systems in presence of crystal-field splitting or difference between the bandwidths.
We will in particularly highlight how in these systems the bands can show very different behaviors when $J$ is sizable. For instance the bands can show a different degree of correlation, and the Mott transition can become selective.

\begin{figure}[htbp]
\begin{center}
\includegraphics[width=8cm]{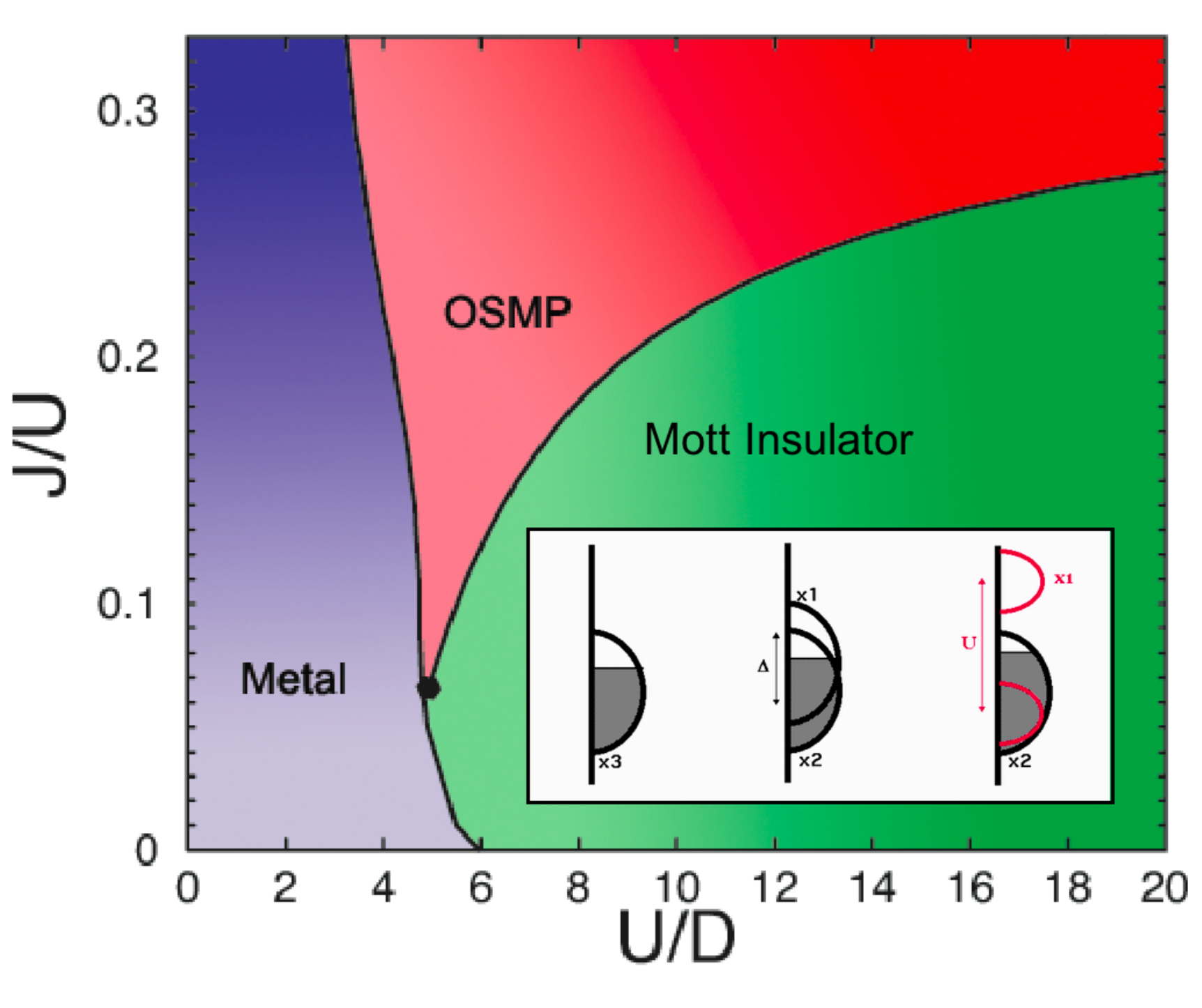}
\end{center}
\caption{Phase diagram from Ref. \cite{demedici_3bandOSMT} for a 3-band Hubbard model populated by 4 electrons, with two degenerate bands and one lifted in energy as depicted in the inset.}
\label{fig:demedici_OSMT}
\end{figure}

OSMT has been thoroughly studied in the half-filled Hubbard model with two different bandwidths\cite{Anisimov_OSMT,Koga_OSMT,demedici_Slave-spins,Ferrero_OSMT,Liebsch_OSMT_3,biermann_nfl,Werner_Hund,Vojta_OSMT_review} and it has been shown how Hund's coupling widens the parameter region where the OSMT can be found. 
However it has been shown in a recent publication\cite{demedici_3bandOSMT} that even systems in which all bands have the same bandwidth can undergo an OSMT when J is beyond a critical value. This has been shown for a 3-band model populated by 4 electrons, in which two bands are degenerate and one is lifted up in energy (inset of Fig. \ref{fig:demedici_OSMT}). For a large range of crystal-field splitting the upper band can open a Mott gap becoming half-filled, while the other two bands remain metallic.

It is worth stressing the important role played by the Hund's coupling. As said, a finite value of J is needed in order to trigger the OSMT. Then further raising $J/U$ widens enormously the selective phase (see Fig. \ref{fig:demedici_OSMT}), pushing the boundary of the Mott insulating phase at very high $U$. This being a localization of 4 electrons in 3 bands, the quick raise of the $U_c$ necessary for this system to become a Mott insulator is explained by our study of the non half-filled Mott transition performed in section \ref{sec:deg}.

What is remarkable here is that while the boundary for the Mott transition of the whole system follows closely the result for the degenerate system (the result for N=3 n=2, equivalent to n=4, of Fig. \ref{fig:Uc_vs_ratioJU}) one band has a completely different behavior and becomes selectively localized at a much smaller critical value of U, which gets further reduced by J.

Even if this effect has been shown not very sensitive to a small splitting of the two degenerate bands, the main effect highlighted in Ref. \cite{demedici_3bandOSMT} as possibly responsible for the OSMT behavior is the different degeneracy of the two parts of the system. Albeit having the same bandwidth, the doubly degenerate subsystem has an enhanced kinetic energy that can be seen as an increased effective bandwidth (thus having a different $U_c$, if it were decoupled from the third band), tracing this OSMT back to the previously studied cases of systems of bands of unequal width.

We will show here that the main effect causing the OSMT in these system of equal bands is another, namely the suppression of the orbital correlations induced by J.

\begin{figure}[htbp]
\begin{center}
\includegraphics[width=8cm]{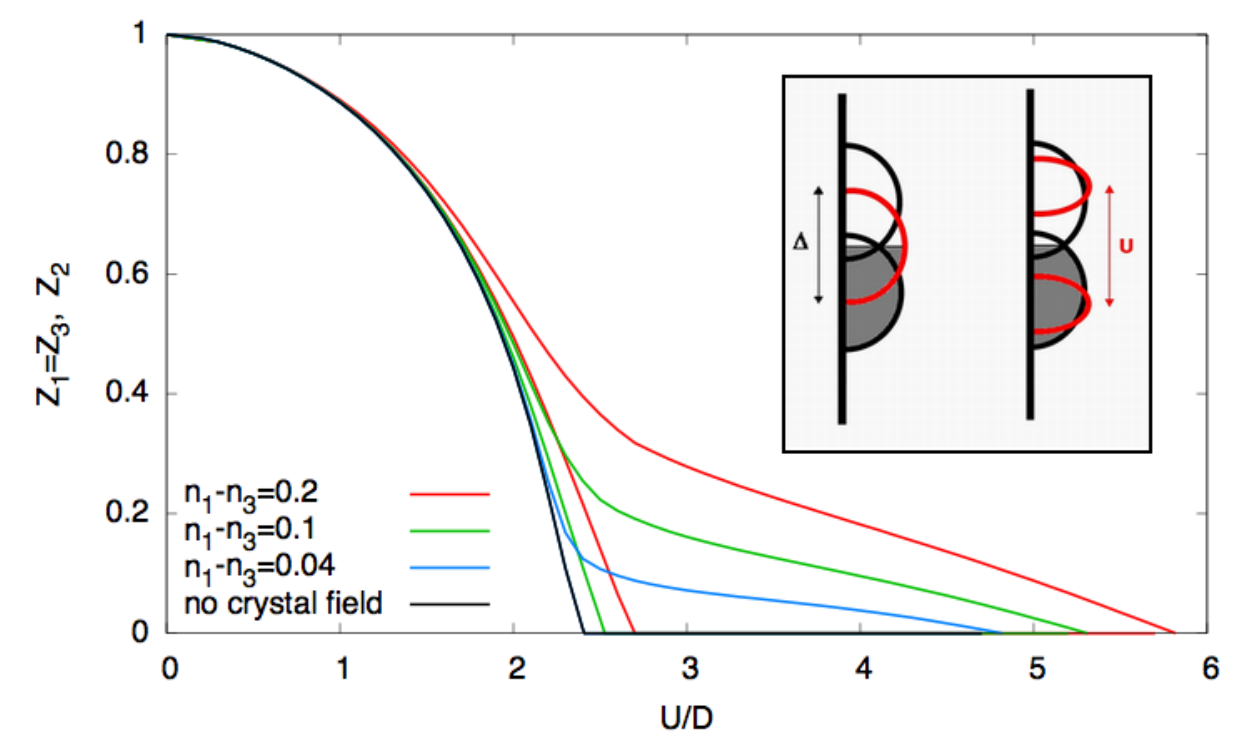}
\end{center}
\caption{SSMF results for a half-filled 3-band Hubbard model with trigonal crystal-field splitting. One band remains half-filled, while the other two are shifted symmetrically, as depicted in the inset. The quasiparticle weights $Z_1=Z_2$,$Z_3$ are shown here, as a function of the interaction strength $U/D$ for Hund's coupling $J/U=0.25$ at different fixed population unbalance $n_1-n_3$.}
\label{fig:1up1down}
\end{figure}
For this we will study a system far from any degeneracy. We consider a 3-band model eq.(\ref{eq:Ham}) with all equal bandwidths and a trigonal crystal-field splitting ($\D_{12}=-\D_{23}$), at half-filling. Namely one band remains half-filled, one is lifted in energy, and another is symmetrically lowered (inset of Fig. \ref{fig:1up1down}). 

\begin{figure}[htbp]
\begin{center}
\includegraphics[width=8cm]{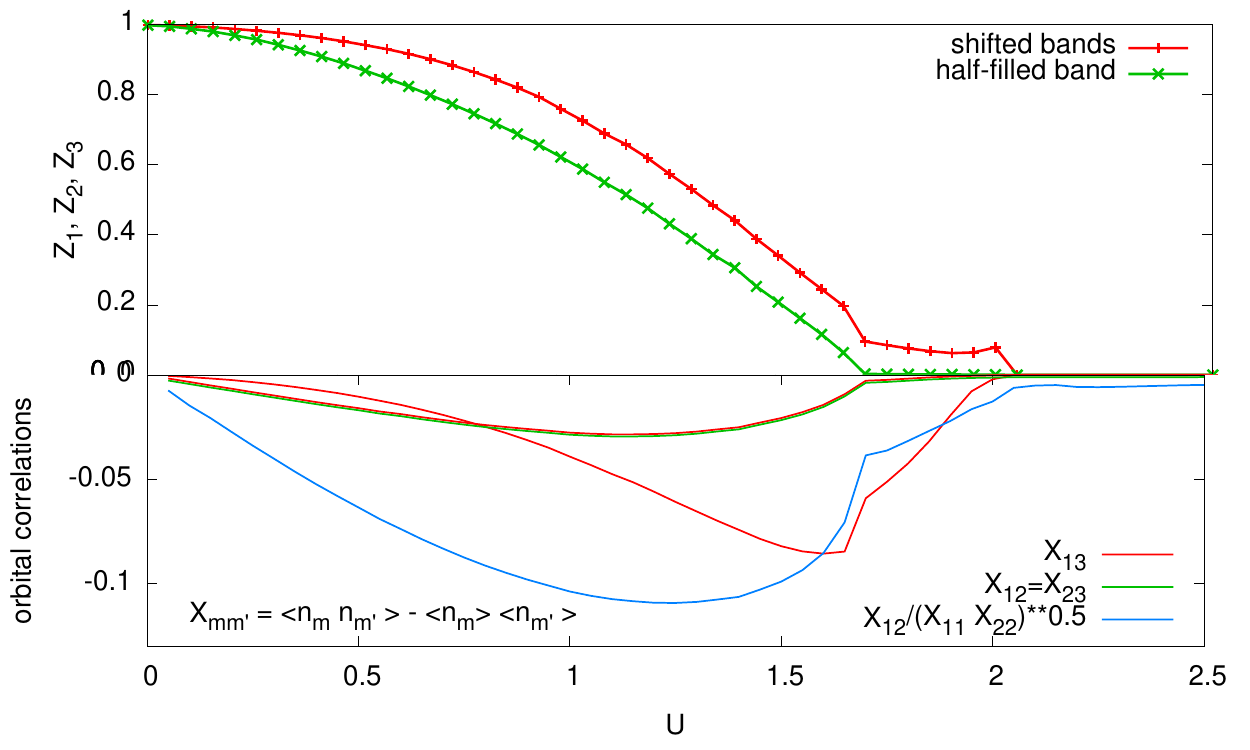}
\end{center}
\caption{DMFT results for the same model as in the inset of Fig. \ref{fig:1up1down}, but fixing the crystal-field splitting $\D_{12}=-\D_{23}=0.7D$. Upper panel: quasiparticle residues $Z_m$. Lower Panel: orbital correlations. A clear suppression of the orbital correlations between the half-filled band and the lifted ones corresponds to the OSMT.}
\label{fig:1up1down_DMFT}
\end{figure}

Results in SSMF for a sizable Hund's coupling $J/U=0.25$ are shown in Fig. \ref{fig:1up1down} for different fixed population unbalance $n_1-n_3$. One finds that the half-filled band becomes insulating at a lower critical coupling than the other two, thus opening an Orbital-Selective Mott Phase. The reason for studying the model at fixed $n_1-n_3$ rather than at fixed $\D$ is that correlations tend to renormalize and strongly reduce the crystal-field splitting, so that keeping the population unbalance fixed the OSMT is enhanced and shown to be possible. We then check the robustness of the OSMT at fixed $\D_{12}=0.7D$ in DMFT and find it in a reduced but still sizable range of correlation strength (Fig. \ref{fig:1up1down_DMFT}). 

A complete study of the phase diagram of this model is beyond the scope of this paper, however we have shown that degeneracy is not necessary for an OSMT to happen in a system of all equal bandwidths. 

What differentiates the bands here is only the respective individual filling. The system as a whole becomes insulating at a critical U very close to the value $U_c=U_c(N=3,n=3,J=0.25)+\D$, where $U_c(N=3,n=3,J=0.25)\simeq1.15D$, consistent with the atomic picture of the Mott insulator in which the high-spin state is still the ground state and the Mott Gap is reduced by the crystal-field splitting.

\begin{figure}[htbp]
\begin{center}
\includegraphics[width=8cm]{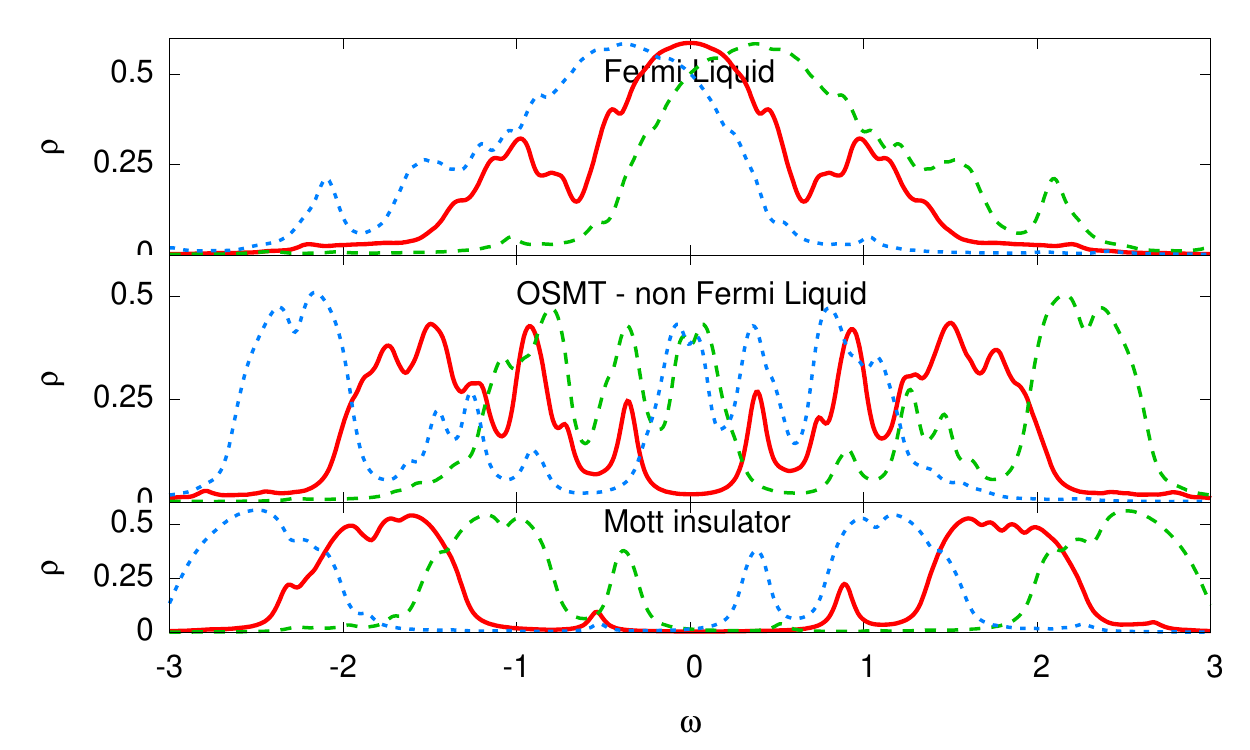}
\end{center}
\caption{Spectral functions corresponding to the 3 phases of the data plotted in Fig. \ref{fig:1up1down_DMFT},  for $J=0.25U$ and $\D_{12}=-\D_{23}=0.7D$. Panels from top to bottom: $U/D=1.0, 1.8, 2.3$. The half-filled bands is indicated in red full line, the two shifted bands in blue and green dashed lines. In the top panel all the bands are metallic, in the bottom panel all are Mott insulating. In the middle panel only the half-filled band opens a Mott gap, whereas the two shifted bands are metallic, but the spectral density at the chemical potential violates the pinning condition imposed from the Luttinger theorem, signaling a non-Fermi-liquid state.}
\label{fig:1up1down_DMFT_spectral}
\end{figure}

However the half-filled band acts as "decoupled" and the value of U at which it becomes insulating is smaller than $U_c$ for the whole system.

The explanation of this behavior becomes transparent when calculating the orbital correlations
\bea
X_{m m^\'}\# & = \# &  \langle (n_m - \langle n_m\rangle)(n_{m^\'}-\langle n_{m^\'}\rangle\rangle  \nonumber \\
\#&=\#&  \langle n_m n_{m^\'}\rangle-\langle n_m\rangle\langle n_{m^\'}\rangle,
\eea
in DMFT (lower panel of Fig. \ref{fig:1up1down_DMFT}). It is clear that the OSMT is accompanied by an almost complete suppression of the orbital fluctuations between the half-filled band and the lifted ones. This implies that close to the transition the half-filled band is in practice decoupled and it can become Mott insulating while the others remain metallic. Also the effect of the crystal field will be minor on its Mott gap. Consequently the $U_c$ for this band will be poorly affected by it.

One may think that this suppression of the \emph{inter}-orbital fluctuations reflects only the suppression of \emph{intra}-orbital fluctuations typical of any Mott transition (owing to the fact that a Mott insulator has a very small double occupancy). This is not the case, here. Indeed we plot in Fig. \ref{fig:1up1down_DMFT} the ratio between these quantities $X_{12}/\sqrt{X_{11}X_{22}}$, i.e. the inter-orbital fluctuations normalized by the standard deviation of the intra-orbital ones in each of the involved bands,  which drops very fast at the OSMT. This shows that the inter-orbital fluctuations are suppressed much faster than the intra-orbital ones and highlights the role of this suppression driven by J in decoupling the bands and thus inducing the OSMT.

In Fig. \ref{fig:1up1down_DMFT_spectral} we show the spectral functions corresponding to the three phases (metallic, orbitally-selective and Mott insulating) that the model goes through. The orbitally-selective Mott phase shows a clear gap in the half-filled band only. It is also characterized by non Fermi-liquid behavior, here represented by a lowering of the density of states at the Fermi level in the metallic bands\footnote{Actually it has been shown\cite{biermann_nfl} that the fully rotational-invariant $H_{int}$ that we study does not lead to a violation of the Luttinger theorem in the orbitally selective phase at strictly zero temperature. The non Fermi-Liquid behavior is rather signaled by a non-linear vanishing of the self-energy at low $\iomn$. This represents an infinite quasiparticle lifetime at the Fermi energy which however is very sensitive to temperature. Accordingly, when using a finite low-energy resolution, as is done by the discrete ED impurity solver (9 total sites) that we used here, this strong temperature dependence shows up as a strong dependence on the low-energy resolution, leading strong violations of the pinning condition of the zero energy value of the spectral function of the metallic bands, as is reported in Fig. \ref{fig:1up1down_DMFT_spectral}.}.

Thus one can conclude that Hund's coupling acts as a band decoupler, leading, to a first approximation, from a collective to an individual behavior for each band. Then it will be each band's structure and filling that determine the correlated properties.

As another paradigmatic example of such a nearly decoupled behavior we have studied a Hubbard model of 3 half-filled bands with different bandwidths and Hund's coupling $J/U=0.25$.  

\begin{figure}[htbp]
\begin{center}
\includegraphics[width=8cm]{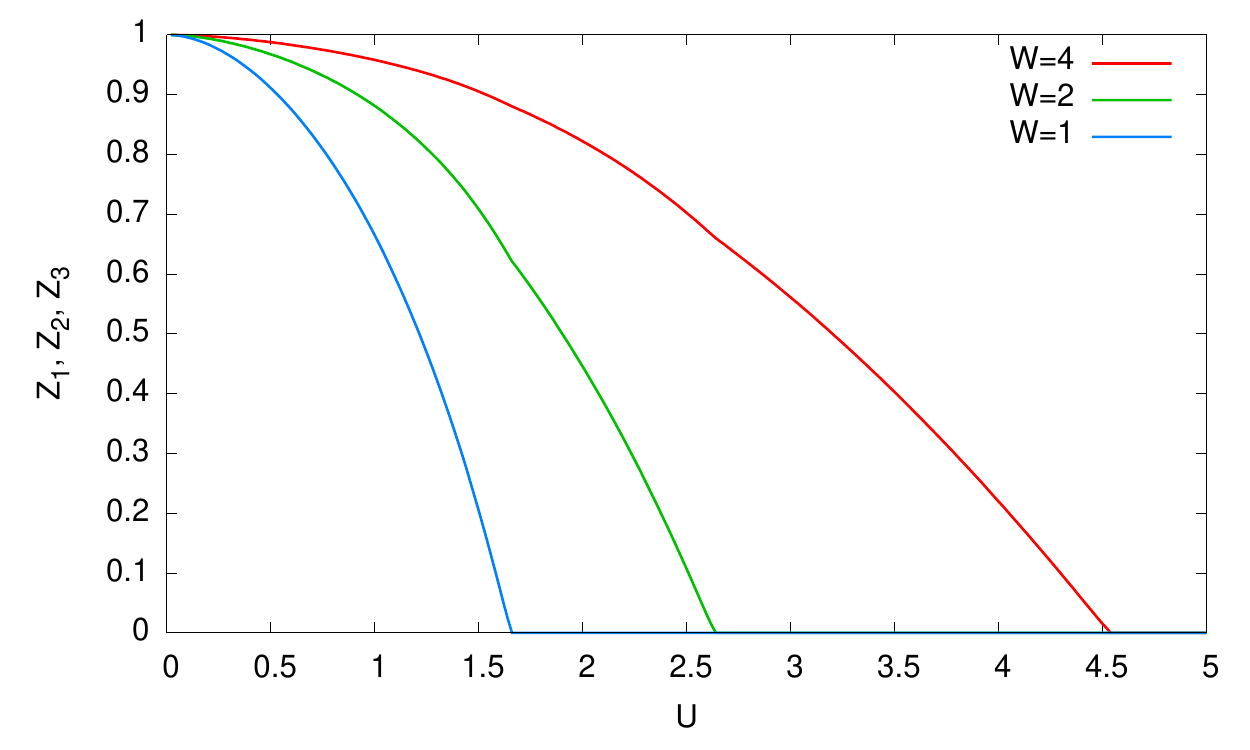}
\end{center}
\caption{SSMF results for a 3-band Hubbard model with bandwidths $W=4,2,1$. All bands are half-filled and $J/U=0.25$. An almost completely decoupled behavior is induced by J, leading to a cascade of OSMT whose critical couplings scale roughly with the bandwidths.}
\label{fig:cascade_SSMF}
\end{figure}

We plot the results in Fig. \ref{fig:cascade_SSMF}. As a function of U the three bands undergo a "cascade" of individual Mott transitions, at $U_c$'s that scale very roughly with the bare bandwidth of each band, thus testifying an almost decoupled behavior.

\begin{figure}[htbp]
\begin{center}
\includegraphics[width=8cm]{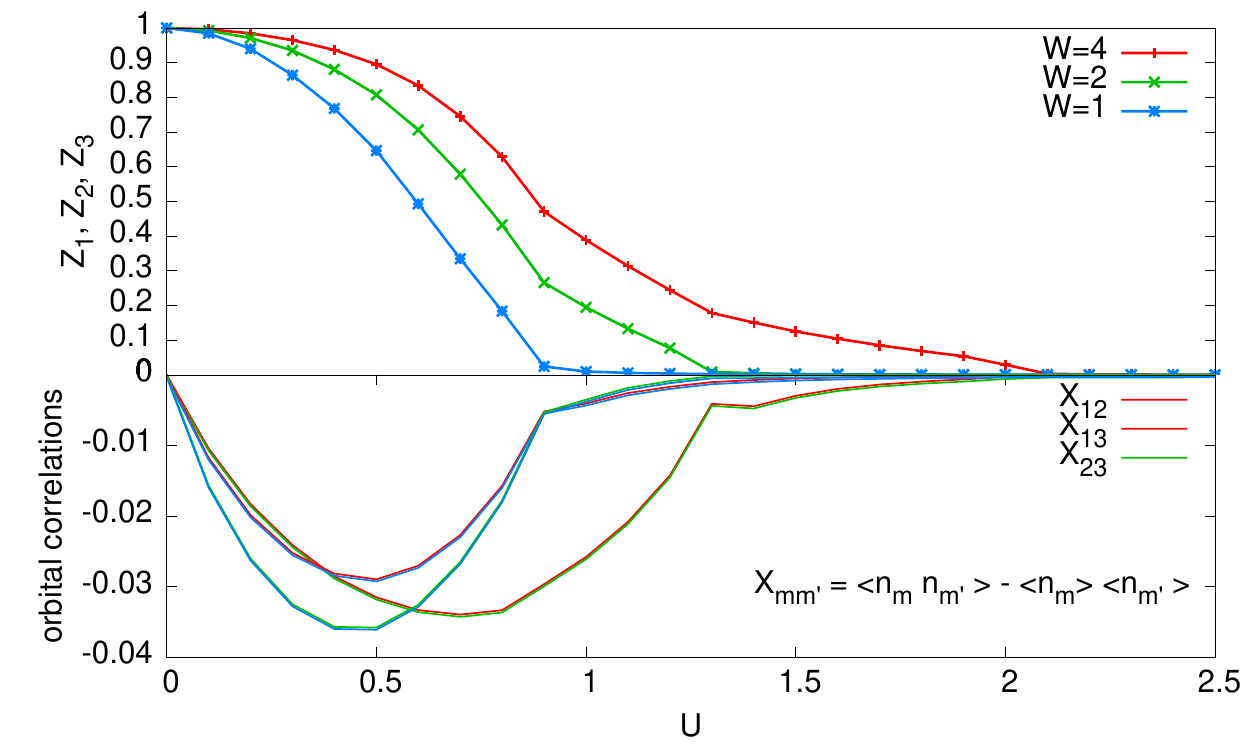}
\end{center}
\caption{DMFT results for the same model as in Fig. \ref{fig:cascade_SSMF}. Upper panel: quasiparticle residue for each band. Lower panel: orbital correlations, which suppressed between the localizing band and the itinerant ones, while they remain sizable between the two itinerant bands.}
\label{fig:cascade_DMFT}
\end{figure}

The SSMF results are confirmed by DMFT as shown in Fig. \ref{fig:cascade_DMFT}, where we also plot the orbital correlations, as previously. The same physics is found here, with a strong suppression of the orbital fluctuations between localized and
itinerant bands, leading to a decoupled behavior.

\section{Conclusions}\label{sec:conclusions}

This work analyzes the role of Hund's coupling in tuning the correlation effects in multi-orbital materials and is divided in two parts.

In the first we have shown that Hund's coupling in degenerate systems of interacting bands can have two opposite effects, depending on the global filling of the system: in half-filled systems it strongly correlates the system and accordingly lowers the critical interaction strength $U_c$ for the Mott transition. In all other cases the $U_c$ is quickly pushed at very high values by J. This is strictly true when J is strong, and it can be fully understood analyzing the atomic multiplets. When J is small instead, the suppression of orbital fluctuations (which is maximal at half-filling and decreases quickly, moving away from it) can dominate, enhancing correlations. This enhances the reduction of $U_c$ in half-filled systems, and produces a reentrant behavior for fillings close to half - before crossing over to the strong J behavior -, while is negligible otherwise.

The second part of this work addresses non-degenerate systems, where crystal field splittings or bandwidth differences characterize the different bands. In these systems we have shown that the Hund's coupling acts mainly as a band-decoupler, through the suppression of the orbital correlations. At intermediate to strong J, the bands will then have different correlation effects, based on the individual structure or filling.
When bands have similar width and structure, the individual proximity to half-filling becomes then a measure of the correlation strength of each band.

In systems where the two effects combine (non-half filling of the global system and proximity to half-filling of one band or a subset of bands)  in presence of sizable Hund's coupling a strong differentiation can be expected, because correlations in the subsystem will be enhanced by J whereas they can be reduced for the rest of the system. 
Strongly and weakly correlated electrons can thus be found in the same conduction bands.\footnote{It is worth stressing that the models studied in this work have zero hybridization between the bands. The conclusions of this work are indeed valid for \emph{weakly hybridized systems}, i.e. they can be extended to nonzero hybridizations with some care.
The notion of weakly hybridized system is a basis-independent one.  Indeed even if a change of basis can always diagonalize the non-interacting band structure, Wannier orbitals in the new basis are not localized, in general and the interaction matrix is very non-local then. We call a "weakly hybridized system" one that can be rotated in a basis where both the interaction is local and the hybridization between the orbitals is small. These two conditions cannot be satisfied for all bandstructures, and are on the contrary a significant constraint, which is material-specific.}

Materials of interest with strong Hund's coupling which arguably fall in this category are i.e. the Iron Superconductors and $Ca_xSr_{2-x}RuO_4$.\cite{Mravlje_SrRuO4_coherence}

\section*{Acknowledgments}
Discussions with Massimo Capone , Lorenzo De Leo, Henri Alloul are gratefully acknowledged. The author is financially supported by Agence Nationale de la Recherche under Program No. ANR-09-RPDOC-019-01 and by RTRA Triangle de la Physique.

\bibliographystyle{apsrev}
\bibliography{Bib/bibldm,Bib/publdm}

\end{document}